\documentclass[aps,preprintnumbers,showpacs,twocolumn,superscriptaddress,floatfix,nofootinbib]{revtex4-1}

\usepackage{latexsym} 
\usepackage{amssymb} 
\usepackage{amsfonts}
\usepackage{amsmath} 
\usepackage{bm} 
\usepackage{mathtools}
\usepackage{envmath}
\usepackage{color}
\usepackage{subfigure} \usepackage{times} \usepackage{units}
\usepackage{hyperref}
\usepackage{lipsum}
\usepackage{float}
\usepackage{latexsym}
\usepackage{envmath}

\def\prl{Phys. Rev. Lett.}
\def\prd{Phys. Rev. D}
\def\cqg{Class. Quantum Grav.}
\def\lrr{Living Rev. Relat.}
\def\apj{Astrophys. J.}

\def\apjl{Astrophys. J. Lett.}
\def\pr{Phys. Rev.}

\def\Dflat{\hat {\mathcal D}}
\def\part_n{\partial_\perp}

\usepackage[utf8x]{inputenc} 
\usepackage{amssymb,amsmath}
\usepackage{graphicx} \usepackage[squaren]{SIunits} 

\begin{document}
\title{Numerical Relativity in Spherical Polar Coordinates: Off-center Simulations}

\author{Thomas W. \surname{Baumgarte}}\affiliation{Department of Physics and Astronomy, Bowdoin College, Brunswick, ME 04011, USA}

\author{Pedro J. \surname{Montero}}\affiliation{Max-Planck-Institute f\"ur Astrophysik, Karl-Schwarzschild-Str. 1, 85748, Garching bei M\"unchen, Germany}

\author{Ewald \surname{M\"uller}}\affiliation{Max-Planck-Institute f\"ur Astrophysik, Karl-Schwarzschild-Str. 1, 85748, Garching bei M\"unchen, Germany}


\begin{abstract}
We have recently presented a new approach for numerical relativity simulations in spherical polar coordinates, both for vacuum and for relativistic hydrodynamics.  Our approach is based on a reference-metric formulation of the BSSN equations, a factoring of all tensor components, as well as a partially implicit Runge-Kutta method, and does not rely on a regularization of the equations, nor does it make any assumptions about the symmetry across the origin.  In order to demonstrate this feature we present here several off-centered simulations, including simulations of single black holes and neutron stars whose center is placed away from the origin of the coordinate system, as well as the asymmetric head-on collision of two black holes.  We also revisit our implementation of relativistic hydrodynamics and demonstrate that a reference-metric formulation of hydrodynamics together with a factoring of all tensor components avoids problems related to the coordinate singularities at the origin and on the axes.   As a particularly demanding test we present results for a shock wave propagating through the origin of the spherical polar coordinate system.
\end{abstract}


\maketitle

\section{Introduction}
\label{sec:intro}

Numerical relativity simulations of black holes and other compact objects have made remarkable progress in recent years.  In particular, simulations of the complete binary black hole coalescence, from inspiral through merger to the quasi-normal ring-down of the merger remnant, became possible with the calculations of \cite{Pre05b,CamLMZ06,BakCCKM06a}.  Since then, a number of different groups have assembled accurate gravitational waveforms emitted in these mergers (see, e.g., the compilation by the NINJA collaboration, \cite{NINJA2}), and have explored astrophysical consequences of these mergers, including black hole recoil (e.g.~\cite{CamLZM07a,GonHSBH07,CamLZM07a,LouZ11}) and spin flip (e.g.~\cite{CamLZKM07}).  

Many current numerical relativity codes (in three spatial dimensions) share several features: they adopt the Baumgarte-Shapiro-Shibata-Nakamura (BSSN) formulation of Einstein's equations \cite{NakOK87,ShiN95,BauS98}, use finite-difference methods in Cartesian coordinates, and adopt moving puncture coordinates, i.e.~a combination of 1+log slicing \cite{BonMSS95} and the ``Gamma-driver" condition \cite{AlcBDKPST03} (a notable exception is the SpEC code; see, e.g., \cite{spec}.)  While Cartesian coordinates are well suited for many applications, in particular simulations of binaries, spherical polar coordinates have some desirable properties whenever the object under consideration is close to spherical or axial symmetry.  Specific examples include gravitational collapse, supernova explosions, and accretion disks.

We have therefore developed and implemented a new approach that applies in spherical polar coordinates the numerical methods that have previously proven to be extremely successful in Cartesian coordinates \cite{BauMCM13}.  As we will review in more detail in Section \ref{sec:spc}, this approach relies on three key ingredients: a reference-metric formulation of the BSSN equations \cite{Bro09,Gou12}, factoring out appropriate geometrical factors from tensor components, and using a ''partially implicit" Runge-Kutta (PIRK) method \cite{MonC12,CorCD12,CorCD14}.  The resulting equations are still singular at the origin of the coordinate system and on the polar axis, but all singular terms can be handled analytically, and the PIRK method is stable even in the presence of these singular terms.  Our approach therefore does not rely on a regularization of the equations, and can be used even in the absence of spherical or axi-symmetry.  In \cite{MonBM14} we applied a reference-metric approach to the formulation of relativistic hydrodynamics, and implemented the resulting equations to perform what we believe are the first self-consistent and stable simulations of general relativistic hydrodynamics in dynamical spacetimes in spherical polar coordinates without the need for regularization or symmetry assumptions.  

While we have previously performed and presented a number of different tests, the vast majority of those tests featured a symmetry about the origin, which raises the question whether the stability of our methods hinges on that symmetry.  In this paper we address this issue by presenting a number of new tests for configurations that do not satisfy that symmetry.  Specifically we will consider an off-centered Schwarzschild black hole (Section \ref{subsubsec:Schwarzschild}), an asymmetric head-on collision of two black holes in (Section \ref{subsubsec:headon}), and an off-centered neutron star (Section \ref{subsubsec:TOV}).

The other purpose of this paper is to revisit our implementation of hydrodynamics.  As we will discuss in more detail in Section \ref{subsec:hydro}, we consider here a modification to the implementation that we presented in \cite{MonBM14}.  As a key test we show in Section \ref{subsubsec:shock} results for a shock wave that propagates through the origin of the coordinate system.

We would like to emphasize that the purpose of the simulations presented in this paper is purely to demonstrate a point of principle.  Placing spherical objects away from the origin of a spherical coordinate system clearly defeats the purpose of such coordinate systems from a computational perspective -- but it does provide extremely powerful tests of the properties of our computational methods.    Moreover, applications for which we expect spherical polar coordinates to be useful, for example supernova collapse or accretion onto a black hole, may involve processes in which asymmetries move the center of the central object away from the origin of the coordinate system -- it is therefore important to calibrate the performance of the numerical methods for such off-center configurations.

\section{Numerical Relativity in Spherical Polar Coordinates}
\label{sec:spc}

\subsection{Einstein's Field Equations}
\label{subsec:EFE}

We refer the reader to \cite{BauMCM13} for a detailed discussion of our implementation of Einstein's field equations, based on the BSSN formulation \cite{NakOK87,ShiN95,BauS98}, in spherical polar coordinates.  Here we provide a brief discussion of the main components, namely a reference-metric formulation of the BSSN equations (Section \ref{subsubsec:BSSN}), a factoring of tensor components (Section \ref{subsubsec:factor}), and a partially-implicit Runge-Kutta scheme (Section \ref{subsubsec:PIRK}).  We will also include brief Sections on the gauge conditions used in this paper (Section \ref{subsubsec:gauge}) as well as the numerical implementation (Section \ref{subsubsec:numerics}).

\subsubsection{A Reference-Metric Formulation of the BSSN Equations}
\label{subsubsec:BSSN}

In a 3+1 decomposition of spacetime the spacetime metric $g_{ab}$ is written as
\begin{equation} \label{spacetime}
ds^2 = g_{ab} dx^a dx^b = - \alpha^2 dt^2 + \gamma_{ij} (dx^i + \beta^i dt)(dx^j + \beta^j dt),
\end{equation}
where $\alpha$ is the lapse function, $\beta^i$ the shift vector, and 
\begin{equation}
\gamma_{ab} \equiv g_{ab} + n_a n_b
\end{equation}
the spatial metric (see, e.g., \cite{Alc08,BauS10,Gou12} for textbook introductions).  In terms of the lapse and the shift, the normal vector $n^a$ on each spatial slice can be written as
\begin{equation}
n_a = (- \alpha,0,0,0)\mbox{~~~~~or~~~~~~}n^a = (1/\alpha, - \beta^i/\alpha).
\end{equation}
Here and in the following indices $a$, $b$ \dots run over spacetime indices, while $i$, $j$ \ldots run over spatial indices only; we also use geometrized units with $c = G = 1$ throughout this paper.  

The BSSN formulation \cite{NakOK87,ShiN95,BauS98} of Einstein's equations further adopts a conformal rescaling of the spatial metric,
\begin{equation} \label{conf_trans}
\gamma_{ij} = e^{4 \phi} \bar \gamma_{ij},
\end{equation}
where $\bar \gamma_{ij}$ is the conformally related metric, $e^\phi$ the conformal factor, and where we refer to $\phi$ as the ``conformal exponent".  This decomposition is not unique, as it allows for different choices of the determinant $\bar \gamma$ of the conformally related metric, which then result in different values of $\phi$,
\begin{equation} 
e^{4 \phi} = (\gamma / \bar \gamma)^{1/3}.
\end{equation}
The original BSSN formulation was based on the choice $\bar \gamma = 1$, which simplifies several expressions.  This choice, which also results in the appearance of tensors with non-zero weight, is appropriate in Cartesian coordinates, but not in curvilinear coordinates.  In spherical polar coordinates one might work around this problem by choosing $\bar \gamma = r^4 \sin^2\theta$ instead, but a more elegant and powerful approach is to adopt a reference-metric formulation (see \cite{Bro09,Gou12}; see also \cite{BonGGN04,ShiUF04}).

In a reference-metric formulation we introduce a new reference metric $\hat \gamma_{ij}$, together with its associated connection $\hat \Gamma^i_{jk}$.  Strictly speaking, only a reference connection is needed for this formalism, but for ease of presentation we assume that this reference connection is associated with a reference metric.  In principle, the reference metric could be any metric, but the formalism is most useful for our purposes when this reference metric is chosen to be the flat metric in whatever coordinate system is used -- in our case in spherical polar coordinates.  We then define 
\begin{equation} \label{diff_connection}
\Delta \Gamma^i_{jk} \equiv \bar \Gamma^i_{jk} - \hat \Gamma^i_{jk},
\end{equation}
where $\bar \Gamma^i_{jk}$ are the connection coefficients associated with the conformally related metric $\bar \gamma_{ij}$.  As the difference between two connections, the coefficients $\Delta \Gamma^i_{ij}$ transform as tensors, unlike the connections themselves.  We compute the coefficients $\Delta \Gamma^i_{jk}$ from
\begin{equation} \label{diff_connection_eval}
\Delta \Gamma^i_{jk} = \frac{1}{2} \bar \gamma^{il} ( \Dflat_j \bar \gamma_{kl} + \Dflat_k \bar \gamma_{jl} - \Dflat_l \bar \gamma_{jk} ),
\end{equation}
where $\Dflat$ denotes the covariant derivative associated with the reference metric $\hat \gamma_{ij}$.   We also define the conformal connection functions as
\begin{equation} \label{conn_fcts}
\bar \Lambda^i \equiv \bar \gamma^{jk} \Delta \Gamma^i_{jk},
\end{equation}
but treat these as new independent variables in the equations.

In order to specify the conformal factor $e^\phi$ we specify the time evolution of the determinant of the conformally related metric,
\begin{equation}
\partial_t \bar \gamma = 0,
\end{equation}
which Brown \cite{Bro09} calls the ``Lagrangian" choice.  

Using the above expressions, the BSSN equations for $\bar \gamma_{ij}$, $\phi$ and other curvature quantities can be expressed independently of any particular choice for $\bar \gamma$; see eqs.~(21) in \cite{Bro09} or eqs.~(9) in \cite{BauMCM13}.  Moreover, many of the differential operators can now be expressed in terms of $\Dflat$.  Choosing $\hat \gamma_{ij}$ to be the flat metric in spherical polar coordinates, 
\begin{equation}
\hat \gamma_{ij} = 
\left(
\begin{array}{ccc}
1 & 0 & 0 \\
0 & r^2 & 0 \\
0 & 0 & r^2 \sin^2 \theta
\end{array}
\right),
\end{equation}
we can then express these differential operators analytically in terms of the spherical polar connection coefficients $\hat \Gamma^i_{jk}$.  

\subsubsection{Factoring of Tensor Components}
\label{subsubsec:factor}

Differential equations, when expressed in spherical polar coordinates, often feature singular terms at the origin or on the axis, where
$r$ or $\sin \theta$ vanish.  The advantage of the reference-metric formulation of Section \ref{subsubsec:BSSN} is that it allows us to express these singular terms analytically.  However, if the variables in the differential equation are tensors, then the tensor components may also become singular at the origin or on the axis.  In order to treat these singular terms analytically as well, we factor out appropriate powers of the geometrical factors $r$ and $\sin \theta$ from tensor components.  

We write the conformal connection functions (\ref{conn_fcts}), for example, as
\begin{equation}
\bar \Lambda^i = \left(
\begin{array}{c}
\lambda^r \\
\lambda^\theta/r \\
\lambda^\varphi/(r \sin \theta)
\end{array}
\right),
\end{equation}
and adopt the coefficients $\lambda^r$, $\lambda^\theta$ and $\lambda^\varphi$ , which remain regular in regular spacetimes, as our dynamical variables.  Covariant derivatives of $\bar \Lambda^i$, for example, can then be expressed in terms of the new variables $\lambda^i$ and their derivatives.  As a concrete example we compute
\begin{equation}
\Dflat_\varphi \bar \Lambda^\theta = \partial_\varphi (\lambda^\theta/r) + \Lambda^i \hat \Gamma^\theta_{i\varphi} 
= \frac{1}{r} \left( \partial_\varphi \lambda^\theta - \cos \theta \lambda^\varphi \right),
\end{equation}
where we have used $\hat \Gamma^\theta_{\varphi\varphi} = - \sin \theta \cos \theta$.  A complete list of all these derivatives is given in eq.~(26) of \cite{BauMCM13}.  As advertised, the singular behavior in the tensor components can now be treated analytically.

We similarly express the conformally related metric as
\begin{equation}
\bar \gamma_{ij} = \hat \gamma_{ij} + \epsilon_{ij},
\end{equation}
where the corrections $\epsilon_{ij}$ do {\em not} need to be small, and then write
\begin{equation}
\epsilon_{ij} = 
\left( \begin{array}{ccc}
h_{rr} 				&  r h_{r\theta}    			& r \sin \theta h_{r\varphi} \\
r h_{r\theta} 			&  r^2 h_{\theta\theta} 		& r^2 \sin \theta h_{\theta \varphi} \\
r \sin \theta h_{r\varphi}  	& r^2 \sin \theta h_{\theta \varphi} & r^2 \sin^2 \theta h_{\varphi\varphi}
\end{array}
\right).
\end{equation}
Similar to our example above, the derivatives $\Dflat_i \bar \gamma_{jk} = \Dflat_i \epsilon_{jk}$ can then be written in terms of the variables $h_{ij}$ -- see eq.~(25) in \cite{BauMCM13} for a complete list.  All other tensorial quantities are treated in a similar way.

With the help of these rescalings, all variables remain finite for regular spacetimes even in spherical polar coordinates.  The equations do feature singular terms, but these singular terms are treated completely analytically.  We do not attempt to regularize the equations; instead we adopt a numerical method that can handle these singular but analytical terms.

\subsubsection{Partially Implicit Runge-Kutta}
\label{subsubsec:PIRK}

The ``partially implicit Runge-Kutta" (PIRK) method was introduced in \cite{MonC12} for the BSSN equations in spherical symmetry (see also \cite{CorCD12,CorCD14}).  In particular, it was demonstrated that the PIRK method can handle the singular terms that appear in spherical polar coordinates as long as they are treated analytically.  We refer to the above references, as well as \cite{BauMCM13}, for a more detailed discussion of the PIRK method; here we illustrate the approach for a simple wave equation
\begin{equation} \label{wave_2nd}
- \partial^2_t \Phi + \nabla^2 \Phi = 0.
\end{equation}
We bring this equation into a form that mimics that of the BSSN equations by introducing a new variable $\kappa \equiv - \partial_t \Phi$.  Also assuming spherical symmetry we then rewrite eq.~(\ref{wave_2nd}) as a pair of first-order-in-time equations
\begin{subequations} \label{wave_1st}
\begin{eqnarray} 
\partial_t \Phi & = & - \kappa \label{wave_psi} \\
\partial_t \kappa & = & - \partial_r^2 \Phi - (2/r) \partial_r \Phi. \label{wave_kappa}
\end{eqnarray}
\end{subequations}
We now recognize that the variable in the singular term, i.e.~$\Phi$ in the term $(2/r) \partial_r \Phi$, is evolved with an equation that does not feature any singular terms (eq.~\ref{wave_psi}).  The idea is then to evolve eq.~(\ref{wave_psi}) explicitly, and use the updated values of $\Phi$ when evaluating the singular terms in eq.~(\ref{wave_kappa}).     In a fully implicit scheme all terms on the right-hand side of the equations would be evaluated using values on the new time level, while in the PIRK scheme only part of the variables are evaluated on the new time levels - namely those that appear in singular terms, which are also those that are evolved with a regular equation.  The effect of this is quite dramatic: while a simple explicit finite-difference evolution of eqs.~(\ref{wave_1st}) quickly becomes unstable, this PIRK method can handle the singular term without problems.  The advantage of PIRK over a fully implicit scheme is that it does not require inversion of any matrices; in fact, the computational cost of PIRK is very similar to that of fully explicit methods.  In a further similarity with fully explicit methods, PIRK is stable only as long as the time step is limited by a Courant condition (see eq.~(\ref{courant}) below).

It turns out that the BSSN equations have a structure similar to that of (\ref{wave_1st}), in particular, all variables in singular terms obey regular equations themselves.  We can therefore apply the PIRK method as described above (see \cite{BauMCM13}, including Appendix B, for details.)

\subsubsection{Gauge Conditions}
\label{subsubsec:gauge}

We adopt different versions of ``moving-puncture" coordinates in this paper, i.e.~a combination of 1+log slicing and the ``Gamma-driver" condition.  Specifically, we adopt both a ``non-advective" version 
\begin{equation} \label{1+log_non_ad}
\partial_t \alpha = - 2 \alpha K
\end{equation}
and an ``advective" version 
\begin{equation} \label{1+log_ad}
\partial_t \alpha - \beta^i \partial_i \alpha = - 2 \alpha K
\end{equation}
of 1+log slicing \cite{BonMSS95} as a condition for the lapse function.  Here $K$ is the trace of the extrinsic curvature
\begin{equation}
K_{ij} = - \frac{1}{2 \alpha} \partial_t \gamma_{ij} + D_{(i} \beta_{j)},
\end{equation}
where $D$ is the covariant derivative associated with the spatial metric $\gamma_{ij}$.  We note that for stationary solutions, for which $\partial_t \alpha = 0$, the non-advective condition (\ref{1+log_non_ad}) is consistent with maximal slicing $K = 0$.

We also use different conditions for the shift vector $\beta^i$.  The simplest choice is $\beta^i = 0$, but we also use an ``non-advective" version of the Gamma-driver condition \cite{AlcBDKPST03}
\begin{subequations} \label{Gamma_driver}
\begin{eqnarray} 
\partial_t \beta^i & = & B^i  \\
\partial_t B^i & = & \mu_S \partial_t \bar \Lambda^i,
\end{eqnarray}
\end{subequations}
where $B^i$ is an auxiliary vector, as well as an ``advective" version of a related condition
\begin{equation} \label{Jena}
\partial_t \beta^i - \beta^j \partial_j \beta^i = \mu_S \bar \Lambda^i
\end{equation}
(see, e.g., \cite{ThiBB11}).  We use $\mu_S = 3/4$ in both conditions.

\subsubsection{Numerical Implementation}
\label{subsubsec:numerics}

Details of our numerical finite-difference implementation can be found in \cite{BauMCM13}, but we review some of the key features here.

We adopt a grid in three spatial dimension, using $(N_r, N_\theta, N_\varphi)$ grid points.  The grid is cell-centered, so that no grid points reside at $r = 0$ or $\sin \theta = 0$.   We use fourth-order differencing to evaluate most spatial derivatives (advective terms are differenced with a third-order upwind scheme); this means that we need to pad the numerical grid with two layers of ghost zones.  Except at the outer boundaries, where we impose simple outgoing-wave fall-off conditions, these ghost zones correspond to another zone in the interior grid (see Fig.~1 in \cite{BauMCM13} for an illustration).  A ghost zone with coordinates $\theta_g$ and $\varphi_g$ and a negative radius $r_g = -\Delta r / 2$, for example, where $\Delta r$ is the radial grid spacing, corresponds to the interior zone at $\theta = \pi - \theta_g$, $\varphi = \varphi_g + \pi$ and $r = - r_g = \Delta r /2$.   The ghost zones can therefore be filled by copying function values from the corresponding interior zones.  For tensor components, appropriate parity conditions have to be taken into account, since unit vectors in the ghost zone may point into the opposite direction of those in the corresponding interior zone (see Table I in \cite{BauMCM13}).  

We implement a second-order version of the PIRK method for the time evolution.  The stability of this method requires that the time step be limited by a Courant condition of the form
\begin{equation} \label{courant}
\Delta t < C \Delta_{\rm min}, 
\end{equation}
where $C$ is a Courant factor and $\Delta_{\rm min}$ is the smallest distance between neighboring grid points.  We evaluate this condition using simple coordinate distances, and chose $C = 0.2$ for all simulations in this paper.   It is a well-known disadvantage of spherical polar coordinates that the accumulation of grid points in the vicinity of the origin leads to a severe limit on the time step.  Nevertheless, we have performed all results shown in this paper with a serial implementation using uniform grids.

In \cite{BauMCM13} we have presented several tests of our code, including convergence tests for Teukolsky waves and single black holes.  Because different parts of the code are differenced to different order, the order of convergence depends on which term dominates the error for the variable under consideration. In \cite{HilBWDBMM13} we also used this code to simulate the collapse of non-linear gravitational waves to black holes. 

While our code does not make any symmetry assumptions, all simulations in this paper are axi-symmetric.  We therefore choose the smallest possible number of grid points in the $\varphi$-direction, $N_\varphi = 2$, in all simulations presented here (but we refer to \cite{BauMCM13} for genuinely three-dimensional simulations with $N_\varphi > 2$).

\subsection{Relativistic Hydrodynamics}
\label{subsec:hydro}

We have previously discussed an implementation of relativistic hydrodynamics in spherical polar coordinates in \cite{MonBM14}.   We briefly review our approach here, and also discuss new features of the approach used in this paper.

\subsubsection{A Reference-Metric Formulation of Relativistic Hydrodynamics}
\label{subsubsec:hydro_reference}

The equations of relativistic hydrodynamics are based on the conservation of rest-mass, expressed by the continuity equation 
\begin{equation}
\nabla_a(\rho_0 u^a) = 0,
\end{equation}
and conservation of energy-momentum
\begin{equation}
\nabla_b T^{ab} = 0.
\end{equation}
Here $\nabla$ denotes the (four-dimensional) covariant derivative associated with the spacetime metric $g_{ab}$, $\rho_0$ the rest-mass density, $u^a$ the fluid four-velocity, and $T^{ab}$ is the stress-energy tensor
\begin{equation}
T^{ab} = \rho_0 h u^a u^b + p g^{ab},
\end{equation}
where $h \equiv 1 + \epsilon + p/\rho_0$ is the enthalpy, $p$ the pressure, and where $\epsilon$ is the specific internal energy.   We also define the Lorentz factor between the fluid and normal observers as
\begin{equation}
W \equiv - n_a u^a = \alpha u^t.
\end{equation}
The quantities $\rho_0$, $p$, $\epsilon$, and the fluid velocity $v^i$, defined as
\begin{equation} \label{v}
v^a \equiv \frac{1}{W} \gamma^a{}_b u^b = \left( 0, u^i/W + \beta^i/\alpha \right),
\end{equation}
form the so-called {\em primitive} variables.\footnote{We note a typo in eq.~(20) of \cite{MonBM14}, which should be replaced with eq.~(\ref{v}) above.}

In many recent applications the above equations are brought into a flux-conservative form, so that high-resolution shock-capturing (HRSC) schemes can be used to find accurate numerical solutions.  In the process, a new set of hydrodynamical variables, namely the {\em conserved} variables, are introduced.   A particularly common such formulation is the so-called ``Valencia" form \cite{BanFIMM97} (see also \cite{MarM03,Fon00} for reviews).

In the Valencia formulation, the continuity equation takes the form
\begin{equation} \label{continuity}
\partial_t (e^{6 \phi} \sqrt{\bar \gamma} D) + \partial_j (f_D)^j = 0,
\end{equation}
the Euler equation is 
\begin{equation} \label{euler}
\partial_t (e^{6 \phi} \sqrt{\bar \gamma} S_i) + \partial_j (f_S)_i{}^j = (s_S)_i,
\end{equation}
and the energy equation becomes
\begin{equation} \label{energy}
\partial_t (e^{6 \phi} \sqrt{\bar \gamma} \tau) + \partial_j (f_\tau)^j = s_\tau.
\end{equation}
Here
\begin{subequations}
\begin{eqnarray}
D & \equiv &  W \rho_0 \\
S_i & \equiv &  W^2 h \rho_0 v_i \\
\tau & \equiv & W^2 h \rho_0 - p - D
\end{eqnarray}
\end{subequations}
are the density, momentum density and the internal energy as seen by a normal observer, 
\begin{subequations}  \label{fluxes}
\begin{eqnarray}
(f_D)^j & \equiv & \alpha e^{6 \phi} \sqrt{\bar \gamma} D (v^j - \beta^j/\alpha) \\
(f_S)_i{}^j & \equiv & \alpha e^{6 \phi} \sqrt{\bar \gamma} (W^2 h \rho_0 h v_i(v^j - \beta^j/\alpha) + p \delta_i{}^j) \\
(f_\tau)^j & \equiv & \alpha e^{6 \phi} \sqrt{\bar \gamma} (\tau (v^j - \beta^j/\alpha) + p v^j)
\end{eqnarray}
\end{subequations}
are the corresponding fluxes, and the two source terms are
\begin{subequations} \label{sources}
\begin{eqnarray}
(s_S)_i & \equiv & \alpha e^{6 \phi} \sqrt{\bar \gamma} ( - T^{00} \alpha \partial_i \alpha + T^0{}_k \partial_i \beta^k + \nonumber \\
	& & ~~~~~~~~~~~~~~ t^{jk} \partial_i \gamma_{jk} / 2) \label{S_source} \\
s_\tau & \equiv & \alpha e^{6 \phi} \sqrt{\bar \gamma} (  t^{ij} K_{ij} - (T^{00} \beta^i + T^{0i}) \partial_i \alpha),
\end{eqnarray}
\end{subequations}
where we have abbreviated
\begin{equation}
t^{ij} \equiv T^{00} \beta^i \beta^j + 2 T^{0i} \beta^j + T^{ij}.
\end{equation}

In curvilinear coordinates, the appearance of the determinant $\bar \gamma$ in the above equations -- in particular in the Euler equation -- poses a problem.   Even for flat space in spherical polar coordinates we have $\bar \gamma^{1/2} = r^2 \sin\theta$.  The $\theta$-dependence of this term leads to the appearance of non-zero terms on both sides of the Euler equation (\ref{euler}), even in spherical symmetry.  Analytically these two terms cancel each other, but since in HRSC schemes the flux terms on the left-hand side are evaluated differently from the source terms on the right-hand side, numerical error will prevent a perfect cancellation.  The resulting error breaks spherical symmetry, and can build up very quickly (see \cite{MonBM14} for a more detailed discussion).

In \cite{MonBM14} we suggested a reference-metric approach, analogous to that applied to the BSSN equation in Section \ref{subsubsec:BSSN}, to solve this problem.\footnote{We note that this problem has been recognized before. In general relativistic hydrodynamics this issue has also been addressed by \cite{Call10,NeiC00}} As we derived in \cite{MonBM14}, the resulting equations are very similar to those of the original Valencia formulation above, except that all appearances of $\sqrt{\bar \gamma}$ have to replaced with $\sqrt{\bar \gamma/\hat \gamma}$ (which immediately solves the problem discussed above), and all partial derivatives $\partial$ have to be replaced with covariant derivatives with respect to the reference metric, $\Dflat$.   The continuity equation (\ref{continuity}), for example, becomes
\begin{equation} \label{continuity_ref}
\partial_t (e^{6 \phi} \sqrt{\bar \gamma/\hat \gamma} D) + \Dflat_j (f_D)^j = 0.
\end{equation}
From a computational perspective, the most important advantage of this reference-metric approach is that the geometrical factors $r^2 \sin\theta$ (and similar for other curvilinear coordinate systems) are eliminated from the definition of the fluxes (\ref{fluxes}).  Alternatively one could, of course, work around this problem by simply scaling out these geometrical factors, but the reference-metric approach has other appealing features as well.   All variables are now defined as tensors of weight zero (unlike those in the original Valencia formulation), and the formalism meshes well with the very similar approach used for the BSSN equations.  The covariant derivatives of the spatial metric $\Dflat_i \gamma_{jk} = 4 e^{4 \phi} \bar \gamma_{jk} \partial_i \varphi + e^{4 \phi} \Dflat_i \bar \gamma_{jk}$, for example,  can be evaluated in terms of the derivatives $\Dflat_i \bar \gamma_{jk}$ that have already been computed for the connection coefficients (\ref{diff_connection_eval}).

We evaluate the covariant derivatives $\Dflat$ in the fluid equations by expanding them in terms of connection coefficients; the continuity equation (\ref{continuity_ref}), for example, becomes 
\begin{equation} \label{continuity_ref_ex}
\partial_t (e^{6 \phi} \sqrt{\bar \gamma/\hat \gamma} D) + \partial_j (f_D)^j = -(f_D)^j \hat \Gamma^k_{jk}.
\end{equation}
The appearance of the new terms on the right-hand side are a disadvantage of this approach.  The vanishing of the right-hand side of the continuity equation in the original Valencia-formulation meant that, in an HRSC implementation, the total rest mass
\begin{equation}\label{restmass}
M_0 = \int d^3 x \sqrt{\gamma} \alpha u^t \rho_0 = \int d^3x \sqrt{\gamma} D,
\end{equation}
is conserved {\em exactly}; this is no longer the case in the reference-metric formulation (see Section \ref{subsubsec:TOV} below for a numerical example.)

In \cite{MonBM14} we therefore considered two different implementations, a {\em full} approach, in which we adopted all three hydrodynamic equations in the reference-metric version, and a {\em partial} approach, in which we adopted the Euler equation in the reference-metric version, but left both the continuity and the energy equation in the original Valencia formulation.  In \cite{MonBM14} we found that the advantages of the partial approach outweighed those of the full approach; however, our implementation there did not adopt a factoring of tensor components.

\subsubsection{Factoring of Tensor Components}
\label{subsubsec:hydro_factor}

In this paper we apply to all hydrodynamical variables the same factoring of tensor components that we described in Section \ref{subsubsec:factor}.  The momentum density $S_i$, for example, is written as 
\begin{equation}
S_i = \left(
\begin{array}{c}
s_r \\
r s_\theta \\
r \sin \theta s_\varphi
\end{array}
\right),
\end{equation}
(and similar for all tensorial variables), and the $s_r$, $s_\theta$ and $s_\varphi$ are then evolved as the dynamical variables.  We found that with this rescaling, and using the full approach, i.e.~adopting the reference-metric approach for all hydrodynamical variables, even a shock wave passing through the origin of the coordinate system will not lead to the formation of spikes or other numerical artifacts -- we will present examples in Section \ref{subsec:hydro_sim}.

\subsubsection{Equation of State}
\label{subsubsec:eos}

For the simulations presented in Section \ref{subsec:hydro_sim} we construct the initial data using a polytropic equation of state (EOS)
\begin{equation} \label{polytrope}
P = \kappa \rho_0^\Gamma,
\end{equation}
where the polytropic constant $\kappa$ is a measure of the entropy, and where $\Gamma$ is related to the polytropic index $n$ by $\Gamma = 1 + 1/n$.   The specific internal energy density $\epsilon$ can then be found from the first law of thermodynamics to be
\begin{equation} \label{gamma-law}
\epsilon = \frac{1}{\Gamma - 1} \, \frac{P}{\rho_0}.
\end{equation}

During the dynamical evolution of these initial data we adopt a $\Gamma$-law EOS, meaning that we use eq.~(\ref{gamma-law}) to find $P$ in terms of the conserved variables 
 $\rho_0$ and $\epsilon$.  

For all simulations in Section \ref{subsec:hydro_sim} we will use units in which the polytropic constant $\kappa$ in (\ref{polytrope}) is unity, $\kappa = 1$.  However, all dimensional quantities can be rescaled for any other value of $\kappa$ by recognizing that, in geometrized units, $\kappa^{n/2}$ has units of length.   Density, for example, has units of inverse length squared, again in geometrized units.  The star considered  in Section \ref{subsubsec:TOV} has a central density of $\rho_c = 0.2$ in our units with $\kappa=1$; for any other value of $\kappa$, the central density is then $\kappa^{-n} \rho_c$. 

\subsubsection{Numerical Implementation}
\label{subsubsec:hydro_numerics}

We use a HRSC scheme to solve the equations of relativistic hydrodynamics in the above form (see, e.g.~\cite{Tor99,Lev92} for text-book treatments).  In particular, we have implemented a second-order slope limiter reconstruction scheme, namely the monotonic centered limiter \cite{Van77}, to obtain the left and right states of the primitive hydrodynamic variables at each cell interface.  We have also adopted the Harten-Lax-van-Leer-Einfeld approximate Riemann solver \cite{HarLL83,Ein88}.  We also refer to \cite{DueLSS05b,ThiBB11} for similar treatments and more detailed discussion.

Following common practice we introduce an artificial atmosphere to deal with the vacuum regions of spacetime, which would otherwise create numerical problems.  We follow the prescription of \cite{ThiBB11} we set the density $\rho_0$ to $\rho_{\rm atm} = f_{\rm atm} \mbox{max}(\rho_0)$ wherever it would otherwise be less than a threshold density $\rho_{\rm thres} = f_{\rm thres} \rho_{\rm atm}$, or wherever $\epsilon < \epsilon_{\rm atm}$.  Here we compute $\epsilon_{\rm atm}$ from $\rho_{\rm atm}$ using a polytropic EOS.  For the simulations presented in Section \ref{subsec:hydro_sim} we used $f_{\rm atm} = 10^{-8}$ and $f_{\rm thres} = 10$.

\section{Off-Center Simulations}
\label{sec:simulations}

\subsection{Vacuum}
\label{subsec:vacuum}

\subsubsection{Off-Centered Schwarzschild Black Holes}
\label{subsubsec:Schwarzschild}

\begin{figure}[tb]
\begin{center}
\includegraphics[width=3in]{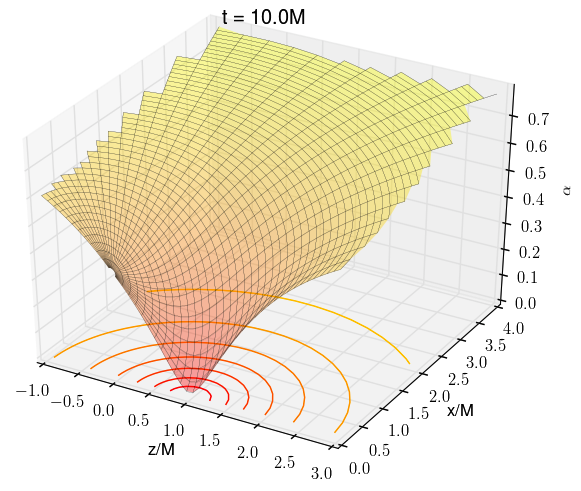}
\end{center}
\caption{The lapse function $\alpha$ in a slice of constant azimuthal angle $\varphi$, evolved on a grid of size $(192,36,2)$.  The color-coded surface shows the data at time $t=10M$, while the wireframe shows the initial data at $t=0$ -- the two surfaces are hardly distinguishable in the figure.  The contour lines are drawn for $\alpha = j \times 0.1$, where $j$ is an integer.}
\label{Fig:trumpet_slice}
\end{figure}

\begin{figure}[tb]
\begin{center}
\includegraphics[width=3in]{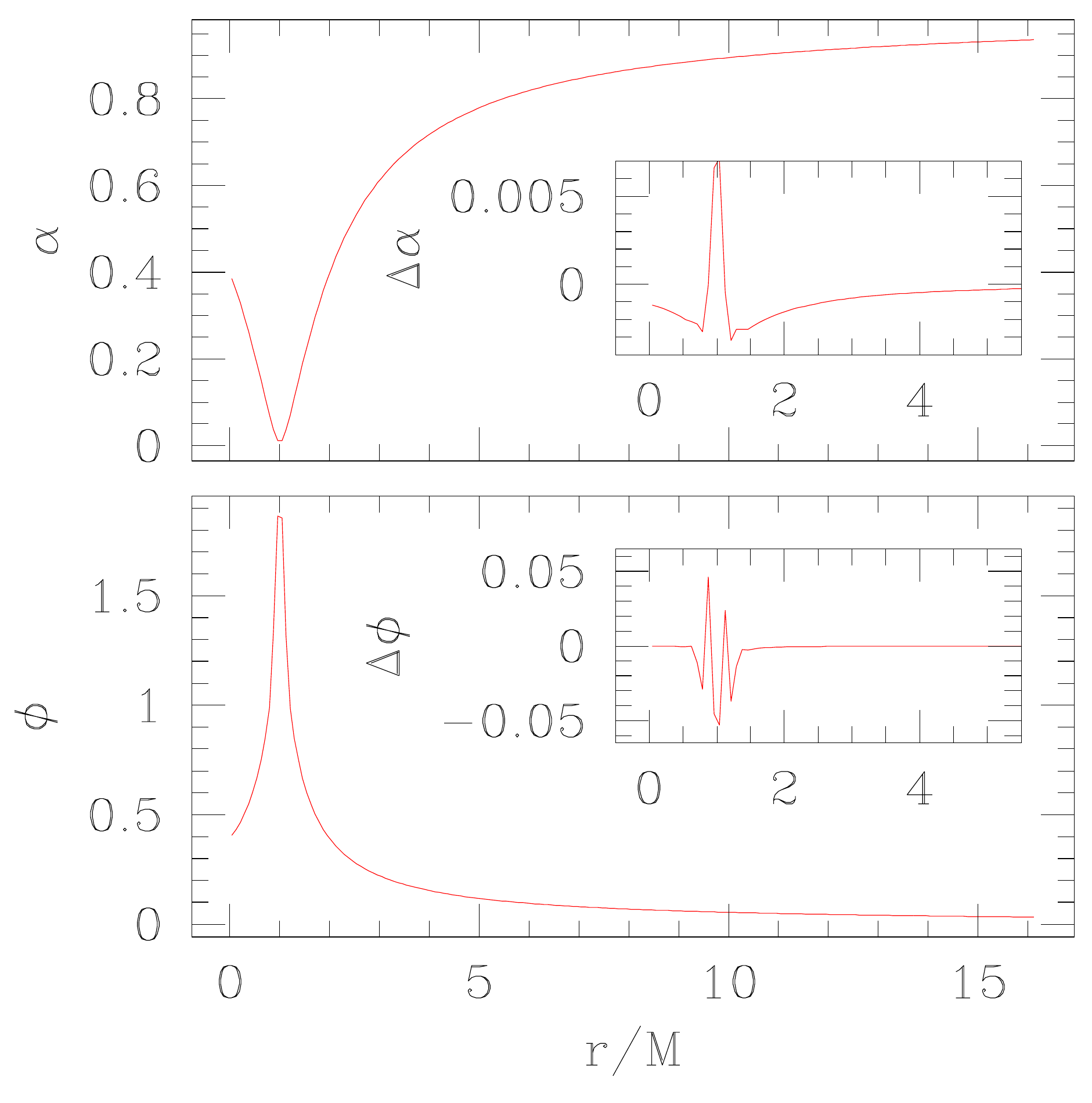}
\end{center}
\caption{The lapse function $\alpha$ (top panel) and the conformal
  exponent $\phi$ (bottom panel) as a function of radius $r$ in the
  $\theta = 0$ direction, at $t =  9.5M$, for the same simulation as
  in Fig.~\ref{Fig:trumpet_slice}.  The insets show the difference of each function between $t = 0$ and $t = 9.5 M$. 
 The center of the black hole is located at $r = 1 M$.}
\label{Fig:trumpet_rays}
\end{figure}

\begin{figure}[tb]
\begin{center}
\includegraphics[width=3in]{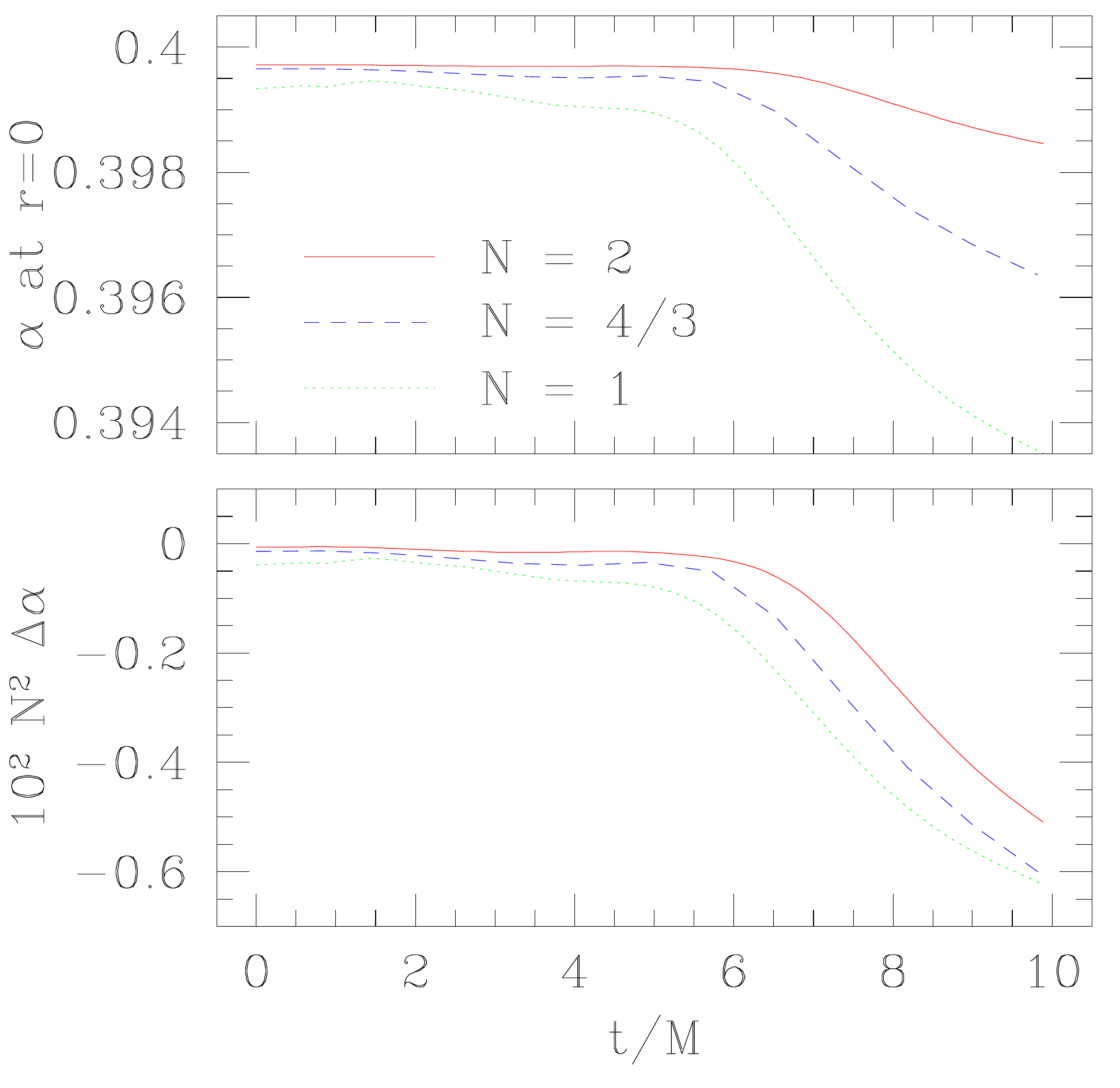}
\end{center}
\caption{In the top panel we show the value of the lapse $\alpha$ at the origin of the coordinate system, $r=0$, as a function of time for a Schwarzschild black hole whose center is located at $r=1M$ and $\theta=0$.   We show results for three different grid resolutions $(N96,N18,2)$ with $N = 1$, $4/3$ and $2$.  In the bottom panel we show the numerical errors $\Delta \alpha$ rescaled with factors of $N^2$, demonstrating at least second-order convergence.}
\label{Fig:trumpet_convergence}
\end{figure}

As a first example of off-centered simulations we consider a Schwarzschild black hole that is placed away from the origin of the coordinate system.  We express this black hole in maximally-sliced ``trumpet" coordinates (see \cite{HanHOBGS06}), which can be expressed analytically in parametric form \cite{BauN07}.   In these coordinates, the black-hole horizon is located at a coordinate distance $r_{\rm hor} = 0.779 M$ away from the black hole's center.  We evolve these data with the non-advective 1+log slicing (\ref{1+log_non_ad}) and Gamma driver condition (\ref{Gamma_driver}), using the analytical values for the lapse and shift as initial data.  Analytically, the resulting evolution should result in all metric components remaining time-independent; any evolution away from the initial data is therefore caused by numerical error.  

We place the center of the black hole at $r_{\rm center} = 1M$ and $\theta = 0$, i.e.~at a distance of $1M$ from the origin of the coordinate system in the positive $z$-direction (the direction towards the North Pole), and choose the numerical grid to extend to an outer boundary imposed at $r_{\rm max} = 16 M$.  We also choose the grid resolution such that the radial resolution $\Delta r = r_{\rm max}/N_r$ is similar to the angular resolution $r \Delta \theta = r \pi/N_{\theta}$ at the center $r_{\rm center}$ of the black hole, which implies
\begin{equation} \label{ang_res}
N_\theta \sim \pi \frac{r_{\rm center}}{r_{\rm max}} N_r.
\end{equation}
In the following we present results for grids of size $(N96,N18,2)$ with $N = 1$, $4/3$ and $2$.  

In Fig.~\ref{Fig:trumpet_slice} we show a ``slice" of the lapse function $\alpha$ at a time $t = 10M$.   Here and in the following, a slice, similar to a slice in an
orange, represents the data as a function of $r$ and $\theta$ for a given value of the azimuthal angle $\varphi$.  Given that initial data for our simulations in this paper are axisymmetric, the particular value of $\varphi$ does not matter.  We then graph the data as functions of Cartesian coordinates $z = r \cos \theta$ and $x = r \sin \theta$.  

In Fig.~\ref{Fig:trumpet_slice} we show the data for our
highest-resolution run with $N=2$ (i.e. on a grid of size
$(192,36,2)$) at time $t=10M$ as the shaded surface.  We also include
the initial data at $t = 0$ as a wireframe representing each grid
point used in this simulation.  As expected, the difference between
the two data sets is very small, so that they cannot be distinguished
in the figure.   In particular, we do not observe any problems arising
at the origin of the coordinate system at $r=0$.  In
Fig.~\ref{Fig:trumpet_rays} we show both the conformal exponent $\phi$
and the lapse function $\alpha$  at $t = 9.5 M$ along the axis
pointing from the origin to the north pole, i.e.~along $\theta =
0$. The insets in Fig.~\ref{Fig:trumpet_rays} display the difference
between the values of these functions at $t = 0$ and $t = 9.5 M$. 

In the top panel of Fig.~\ref{Fig:trumpet_convergence} we show the lapse function $\alpha$, interpolated to the origin at $r=0$, as a function of time, for the three different resolutions $N = 1$, $4/3$ and $2$.  Analytically, the lapse should remain exactly constant; any departure from this constant value is therefore a measure of the numerical error.  In the bottom panel we plot this numerical error $\Delta \alpha \equiv \alpha - \alpha_{\rm ana}$ and multiply the result with $N^2$.  This graph demonstrates that the errors decrease slightly faster than second-order.

Before closing this section we point out that, by placing the black hole away from the origin of the coordinate system, the numerical error becomes asymmetric about the center of the black hole.  The angular resolution, $r \Delta \theta$, at  a given distance away from the center of the black hole, is smaller on the side of the black hole facing the origin of the coordinate system than on the side facing away from the origin (see Fig.~\ref{Fig:trumpet_slice}.)  We have observed that this asymmetric error results in a slow drift of the black hole towards the origin.  This drift cannot yet be seen in Figs.~\ref{Fig:trumpet_slice} and \ref{Fig:trumpet_rays}, but the fact that the lapse starts to decrease at later times in Fig.~\ref{Fig:trumpet_convergence}, as the center of the black hole slowly approaches the origin, is a symptom of that drift.  However, this figure also demonstrates that this drifts converges away as the numerical resolution is increased.

\subsubsection{Head-on collision of two black holes}
\label{subsubsec:headon}

\begin{figure}[t]
\begin{center}
\includegraphics[width=2.6in]{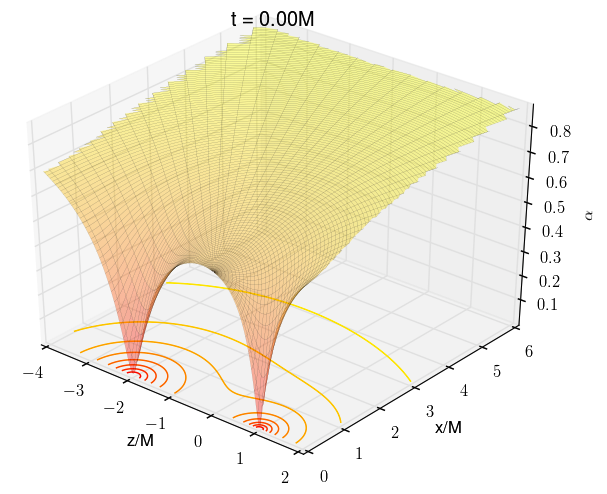}
\includegraphics[width=2.6in]{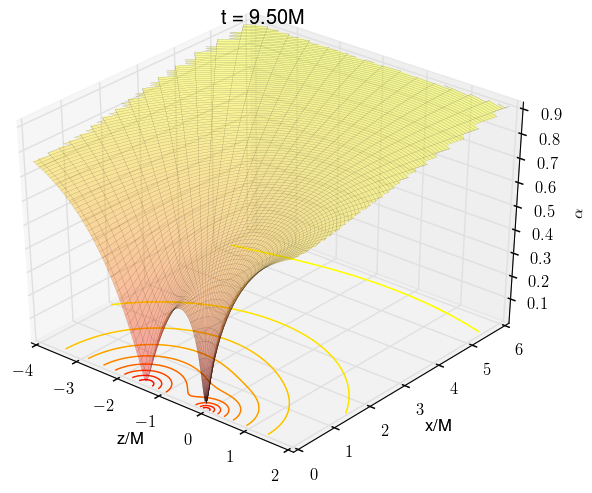}
\includegraphics[width=2.6in]{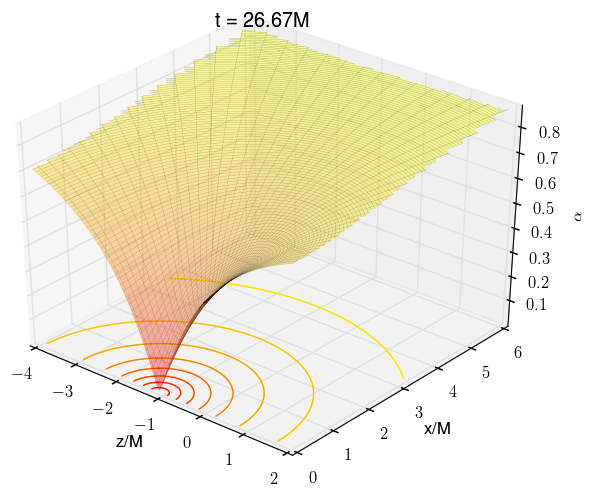}
\end{center}
\caption{Snapshots for the head-on collision of two black holes.  The top panel shows the initial data for the lapse function $\alpha$, based on Brill-Lindquist data; the middle panel at the instant when the smaller black hole moves through the origin of the coordinate system; the bottom panel at a late time after merger.  As in Fig.~\ref{Fig:trumpet_slice}, the contours are drawn for values $\alpha = j \times 0.1$, where $j$ is an integer.  In these plots we show only every second grid point for clarity.}
\label{Fig:BL_snapshots}
\end{figure}

\begin{figure}[t]
\begin{center}
\includegraphics[width=3in]{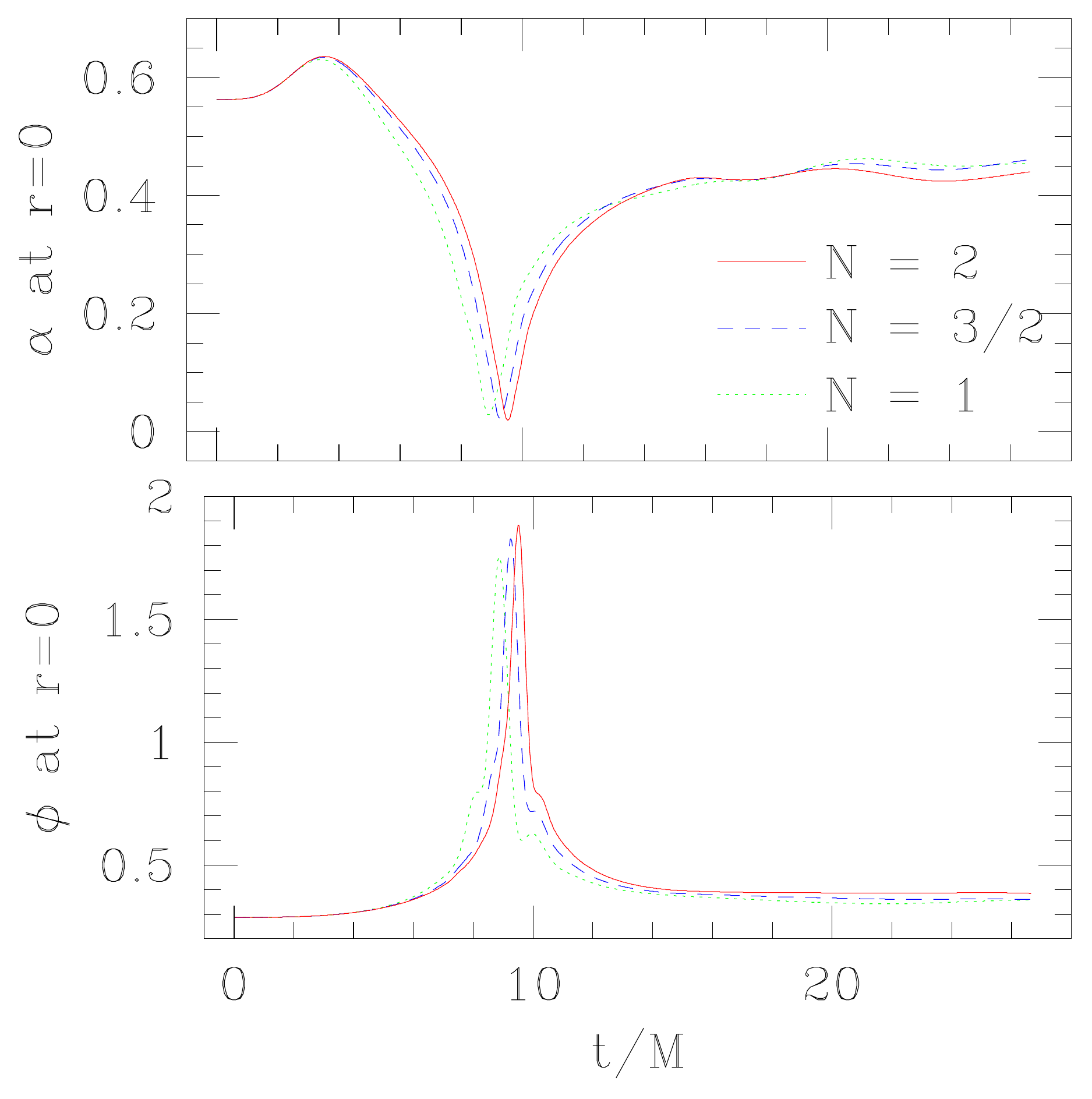}
\end{center}
\caption{ The lapse function $\alpha$ (top panel) and the conformal exponent $\phi$ (bottom panel), evaluated at the origin $r = 0$ of the coordinate system, as a function of time, for the head-on collision of two black holes. }
\label{Fig:BL_lapse}
\end{figure}

As a second example of off-centered simulations in vacuum spacetimes we consider the head-on collision of two black holes from rest.   As initial data we adopt Brill Lindquist \cite{BriL63} data, for which the spatial metric is conformally flat, the extrinsic curvature vanishes, and the conformal factor $\psi = e^\phi$ is given by
\begin{equation} \label{BL}
\psi = 1 + \frac{{\mathcal M}_1}{r_1} + \frac{{\mathcal M}_2}{r_2}.
\end{equation}
Here $r_\alpha \equiv | x^i - C^i_\alpha|$ is the coordinate distance
to the center of the black hole located at $C^i_\alpha$.   We choose both black holes to start out on the axis.  In order to obtain asymmetric data we choose ${\mathcal M}_2 = 2 {\mathcal M}_1$, so that the total ADM mass $M$ of the initial data is $M = 3 {\mathcal M}_1$.  We place ${\mathcal M}_1$ at $z_1 = 3 {\mathcal M}_1 = M$, and ${\mathcal M}_2$ at $z_2 = - 6 {\mathcal M}_1 = - 2 M$.  The center of mass is then at $z_{\rm CM} = - M$.  After the two black holes coalesce, we expect them to merge close to their center of mass (we do not compute the energy or linear momentum emitted in gravitational radiation in these simulations.)  This means that ${\mathcal M}_1$ will pass through the origin of the coordinate system prior to merger, making this a strong test of our numerical methods in spherical polar coordinates.

We evolve these data with the advective version of 1+log slicing, eq.~(\ref{1+log_ad}), as well as the shift condition (\ref{Jena}).  As initial data for these gauge conditions we use the ``pre-collapsed" lapse
\begin{equation}
\alpha = \psi^{-2} = e^{-2 \phi},
\end{equation}
as well as vanishing shift, $\beta^i = 0$.  We also experimented with an advective version of the Gamma-driver condition (\ref{Gamma_driver}), but these simulations crashed shortly after ${\mathcal M}_1$ passed through the origin of the coordinate system.  

For the results presented in this paper we chose a numerical grid extending to an outer boundary at $r_{\rm max} = 24 {\mathcal M}_1 = 8 M$.   Using expression (\ref{ang_res}) we chose the angular resolution similar to the radial resolution at the initial position $r_1 = 3 {\mathcal M}_1$ of ${\mathcal M}_1$.   We present results for the three different grid sizes $(N128,N48,2)$ with $N = 1$, 3/2 and 2.  

In Fig.~\ref{Fig:BL_snapshots} we show slices of the lapse function $\alpha$ at three different times.  The top panel shows the initial data with the two black holes at their initial positions.   The middle panel shows the data at a time of $9.5M$, at the moment that ${\mathcal M}_1$ passes through the origin of the coordinate system.   Finally, the bottom panel shows the remnant at a time well after merger.  The grid lines in these slices represent every second grid point used in the simulations, and we note that the coordinate singularities at the origin of the coordinate system as well as on the axis do not cause any numerical problems.

In Fig.~\ref{Fig:BL_lapse} we show the lapse function $\alpha$ (top panel) and the conformal exponent $\phi$ (bottom panel) interpolated to the origin of the coordinate system as a function of time.    The lapse function initially increases (until $t \sim 4 M$), which is a result of the coordinate transformation from wormhole-type initial data to a trumpet geometry.  It then drops to a value close to zero at $t \sim 9.5M$, when ${\mathcal M}_1$ passes through the origin; the conformal exponent $\phi$ has a sharp maximum at a similar time.  At later times both the lapse and the conformal factor settle down to equilibrium values.  

We include results for the three different resolutions $N = 1$, 3/2 and 2 in Fig.~\ref{Fig:BL_lapse}.   These resolutions are sufficient to establish convergence at early times, before the origin of the coordinate system has become affected by the finite differencing across the singular conformal exponent at the center of black hole ${\mathcal M}_1$.  At later times, however, these resolutions are not yet sufficient to establish convergence, as the error still appears to be dominated by higher-order error terms.  We found a very similar behavior even at early times for only slightly smaller grid resolutions.


\subsection{Relativistic Hydrodynamics}
\label{subsec:hydro_sim}

\subsubsection{Planar Shock-Tube}
\label{subsubsec:shock}

\begin{figure}[t]
\begin{center}
\includegraphics[width=3in]{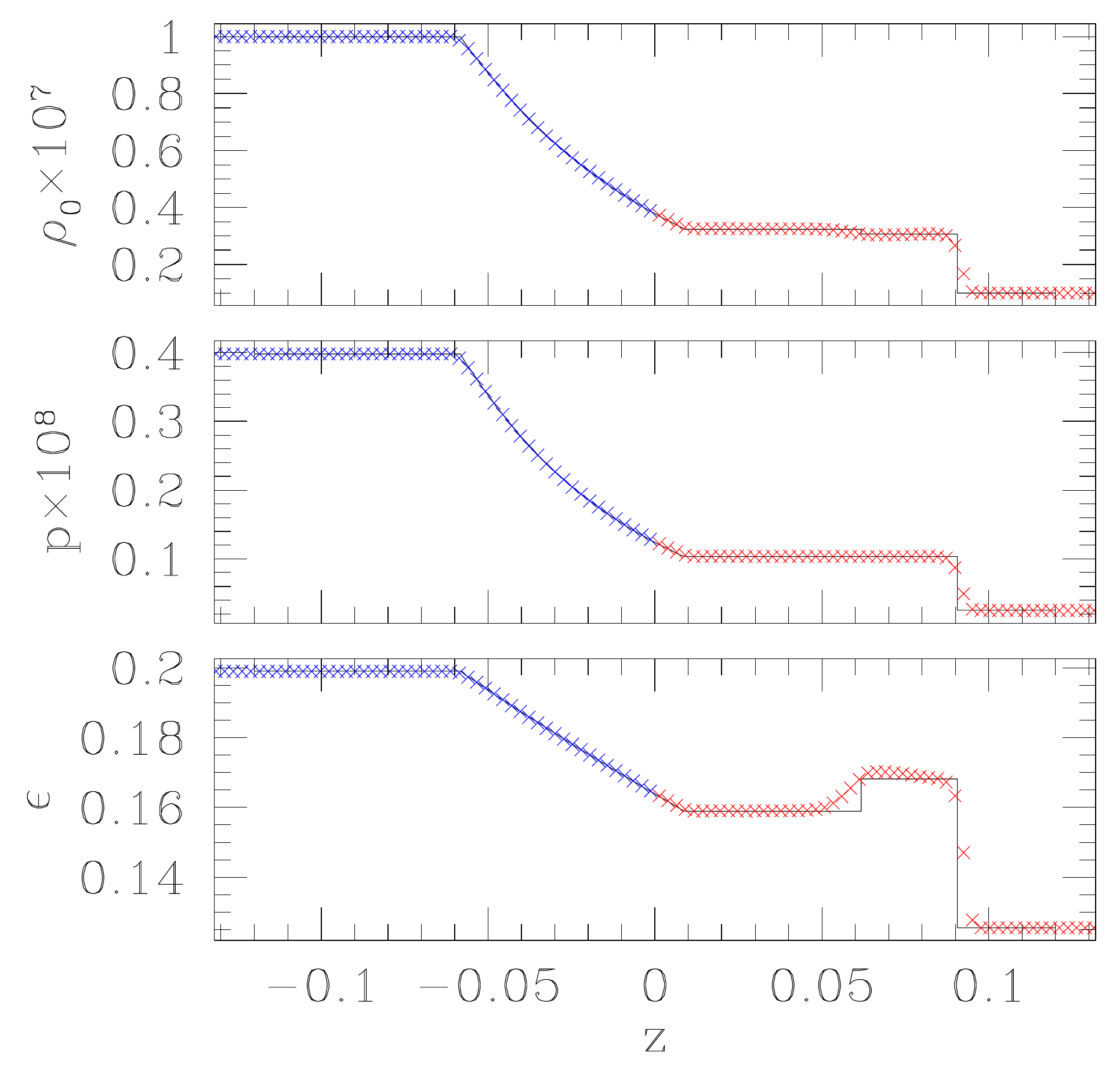}
\end{center}
\caption{The rest-mass density $\rho_0$, the pressure $p$, and the specific internal energy $\epsilon$, interpolated to the axis, for the shock-tube problem described in the text.   The solid line denotes the analytical solution at time $t = 0.3$; the crosses mark our numerical solution.  The two different colors (blue and red) represent data on the northern and southern axis (i.e.~for $\theta = 0$ and $\theta = \pi$);  we do not observe any problem at the origin of the coordinate system, where the two colors meet.}
\label{Fig:shock}
\end{figure}

\begin{figure}[t]
\begin{center}
\includegraphics[width=2.6in]{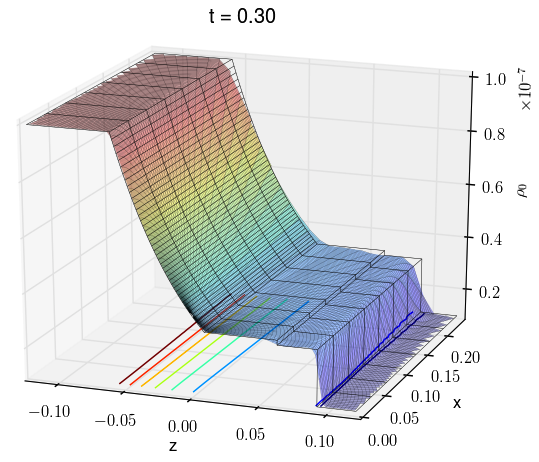}
\includegraphics[width=2.6in]{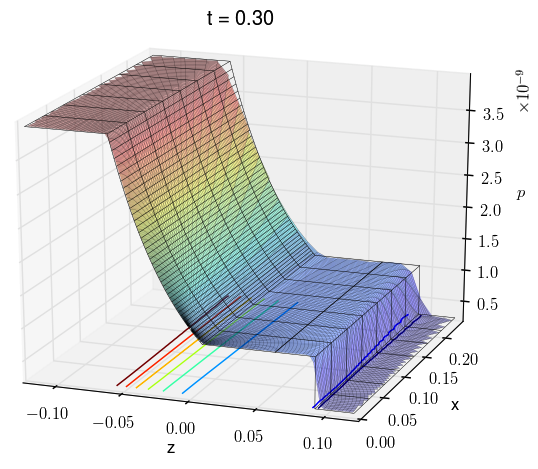}
\includegraphics[width=2.6in]{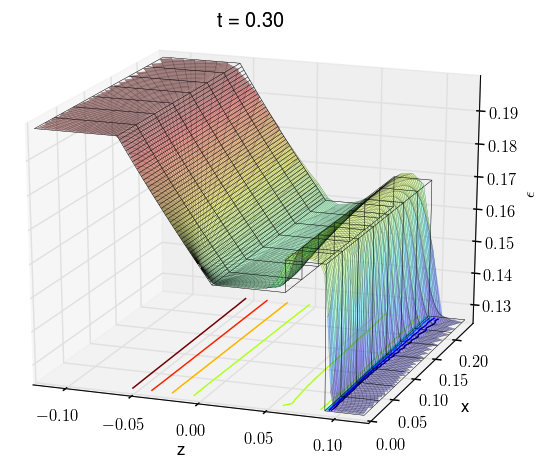}
\end{center}
\caption{The rest-mass density $\rho_0$, the pressure $p$, and the specific internal energy $\epsilon$ on a slice of constant azimuthal  angle $\varphi$ for a planar shock-tube problem.  The color-coded  surface shows the numerical data at time $t=0.3$, while the wireframe shows the analytical data at the same time.   The contour lines are drawn at eight equidistant values between the minimum and maximum values of the respective variables.}
\label{Fig:shock_slices}
\end{figure}

As a test of our implementation of relativistic hydrodynamics in
spherical polar coordinates we first present results for a planar,
special relativistic shock-tube problem.  We consider a fluid at rest
with two different homogeneous densities in the two hemispheres,
separated by a diaphragm in the equatorial plane (i.e.~at $z = 0$) until a time $t = 0$.
At $t=0$ the  partition is removed, which results in a shock that
propagates into the low-density region, while a rarefaction wave
propagates into the high-density region.  The analytical solution for
this special relativistic shock tube problem is given in \cite{Tho86}
(see \cite{MarM03} for the general solution). 

We choose an equation of state as described in Section \ref{subsubsec:eos}, in particular we set up polytropic initial data with a polytropic index $\Gamma = 1.2$, and evolve these data with the Gamma-law EOS (\ref{gamma-law}).  For the example presented here we set 
\begin{equation}
\rho_0 = 
\begin{cases}
10^{-8},& \theta < \pi/2  \\
10^{-7}, & \theta > \pi/2 
\end{cases}
\end{equation}
(in our units with $\kappa = 1$; see Section \ref{subsubsec:eos}).   While we do evolve the spacetime together with the fluid for this test, we have chosen the densities sufficiently small so that the spacetime remains very close to flat.


For this test we imposed an outer boundary at $r_{\rm max} = 0.5$, and chose a numerical grid of size $(192,96,2)$.  We use the ``full" version of our reference-metric formulation of hydrodynamics in this Section, meaning that all hydrodynamical equations are expressed in terms of a reference metric (see Section \ref{subsubsec:hydro_reference}). In the following we show results at time $t = 0.3$, after which the shock has travelled to $z \approx 0.09$.

In Fig.~\ref{Fig:shock} we show the analytical (solid line) and numerical solutions (crosses) at $t=0.3$.   For this graph we interpolated the numerical data to the axis $\theta = 0$, so that they can be compared directly with the analytical solution.  For most parts of the solution the agreement is excellent.  As expected, the shock itself is spread out over about three grid points.  Also as expected, the contact discontinuity (at about $z = 0.06$ in Fig.~\ref{Fig:shock}) poses the greatest numerical challenge, especially in the specific internal energy. However, we have compared with lower-resolution results to confirm that increasing the numerical resolution  results in an improved representation of this discontinuity.  What distinguishes this figure from many other shock-tube tests, however, is the presence of the origin of the coordinate system.  In order to highlight this, we plotted grid points on the northern axis ($\theta = 0$) in red, and those on the southern axis ($\theta = \pi$) as blue.  We do not observe any numerical problems at the origin of the coordinate system, where the two colors meet.

Fig.~\ref{Fig:shock_slices} shows the fluid variables at the same time $t=0.3$, but represented as surface plots.  The colored surfaces show numerical ``slices" of the fluid variables, with the grid lines again representing all grid points used in these simulations.  The (square) wireframe shows the analytical solution.  We again notice very good agreement between the numerical and analytical solutions.  The shock is smeared out across three grid points everywhere; however, since the angular resolution decreases further away from the axis (i.e.~for larger values of $x$), the shock also becomes less sharp.  The contour lines indicate that the entire shock remains very close to planar, even though it is represented in a coordinate system that does not reflect this symmetry.  Most importantly, we again observe that no numerical problems arise at the origin of the coordinate system.   This test therefore demonstrates that our numerical implementation, which includes the reference-metric formulation of hydrodynamics together with a proper factoring of all tensor components, allows for the simulation of a shock wave in spherical polar coordinates, even at the origin of the coordinate system.

We also experimented with planar shocks that originate from a partition located at values of $z < 0$ (i.e.~not in the equatorial plane).  The disadvantage of this setup is that the initial discontinuity is not aligned with cell boundaries of our numerical grid.  This results in a noisy representation of the initial data, which, in turn, results in numerical errors that are larger than those encountered in the experiments described above -- even before the shock wave reaches the origin.  While we did not encounter any instabilities or problems at the origin in these simulations, even after the shock wave had passed through it, we decided to focus on shocks originating from the equatorial plane here, so that the numerical evolution is not affected by numerical noise in the initial data.


\begin{figure}[t]
\begin{center}
\includegraphics[width=3in]{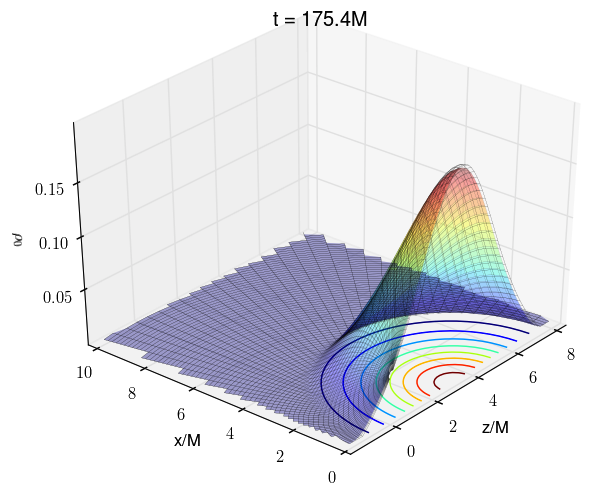}
\end{center}
\caption{The rest-mass density $\rho_0$ on a slice of constant azimuthal angle $\varphi$.  The color-coded surface shows the data at time $t=175.4 M$, while the wireframe shows the initial data at $t=0$.  The contour lines are drawn for eight equidistant densities.}
\label{Fig:TOV}
\end{figure}

\subsubsection{Off-centered Tolman-Oppenheimer-Volkoff Stars}
\label{subsubsec:TOV}

\begin{figure}[t]
\begin{center}
\includegraphics[width=3in]{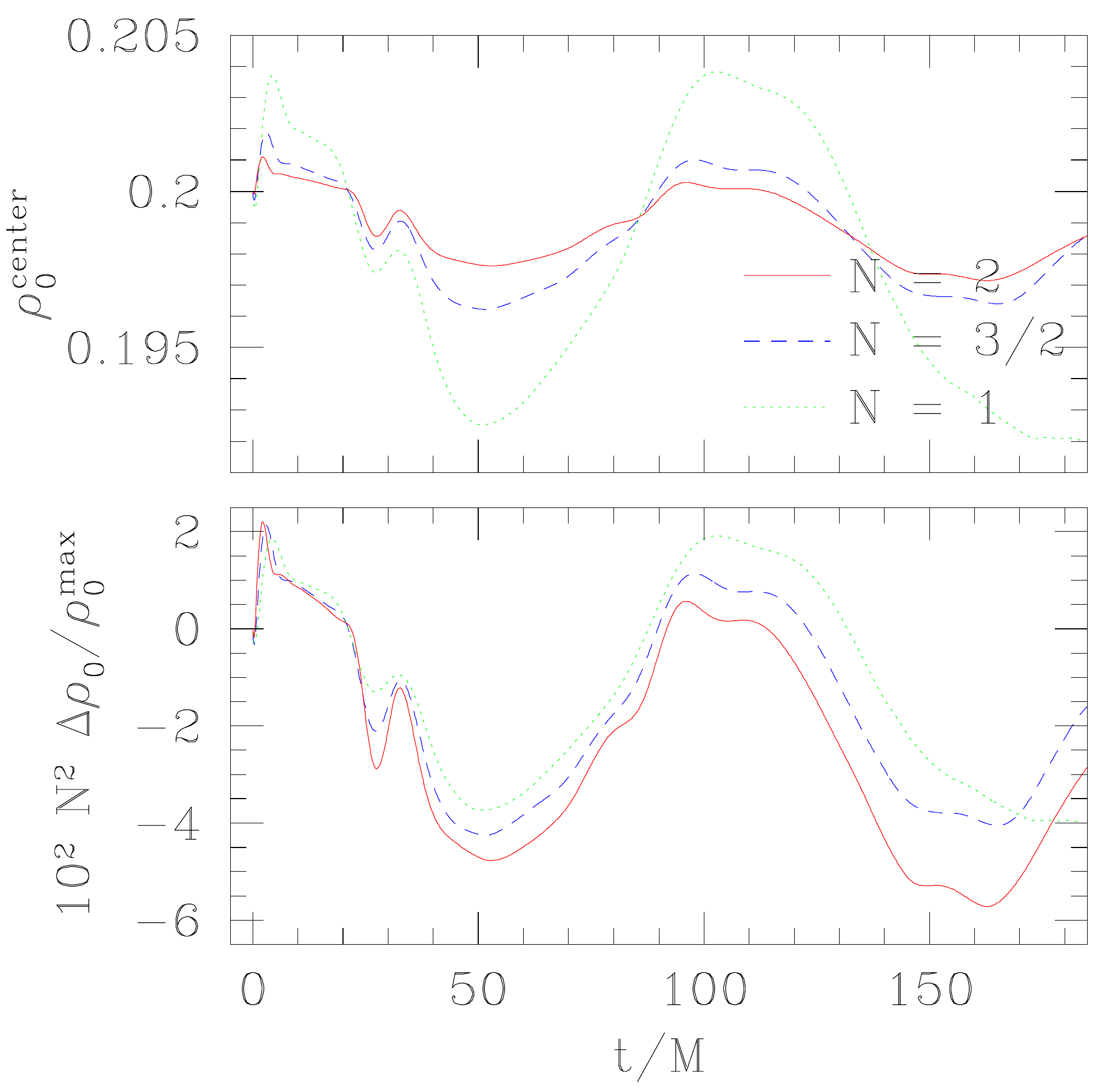}
\end{center}
\caption{The central value of the rest-mass density $\rho_0$, interpolated to $r = 0.5 = 3.18 M$ and $\theta = 0$ for a star with initial maximum density $\rho_0^{\rm max} = 0.2$.  The top panel shows the data themselves for different resolutions; the bottom panel shows the relative errors $\Delta \rho_0/\rho_0^{\rm max}$, with  $\Delta \rho_0 \equiv \rho_0 -  \rho_0^{\rm max}$, rescaled with a factor $N^2$.   We observe second-order convergence until $ t \sim 25 M$, at which point the center of the star is affected by errors originating from the surface of the star.  The simulations converge even after that time, but at a rate slightly less than second-order. }
\label{Fig:TOV_density}
\end{figure}

\begin{figure}[t]
\begin{center}
\includegraphics[width=3in]{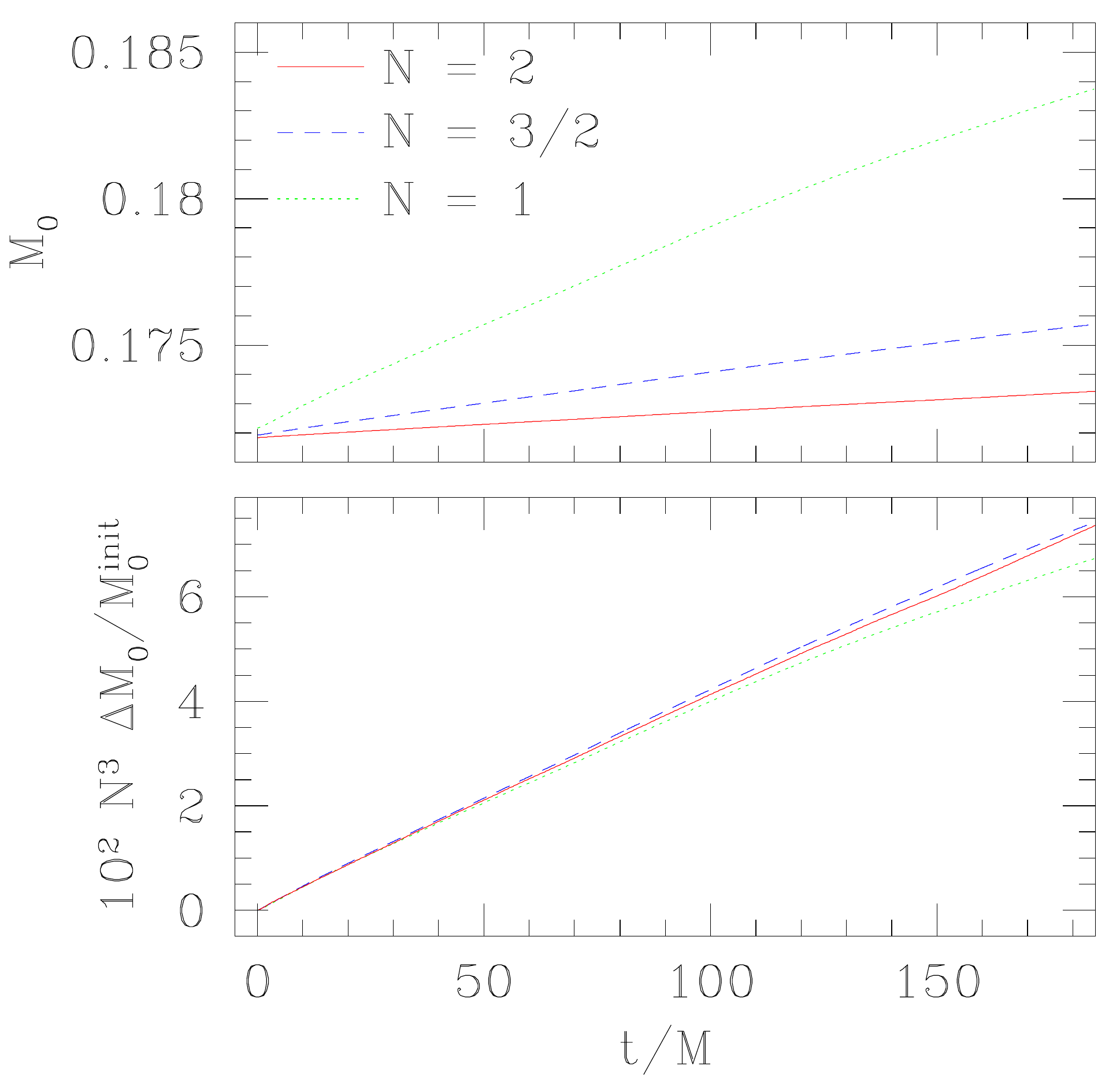}
\end{center}
\caption{The rest mass $M_0$ as a function of time for the same simulations as Fig.~\ref{Fig:TOV_density}.  The top panel shows the data for different resolutions; the bottom panel shows the relative errors $\Delta M_0/M_0^{\rm ana}$, with  $\Delta M_0 \equiv M_0 - M_0^{\rm ana}$, rescaled with a factor $N^3$.  We find that the rest mass converges to approximately third order.}
\label{Fig:TOV_mass}
\end{figure}

\begin{figure}[t].
\begin{center}
\includegraphics[width=3in]{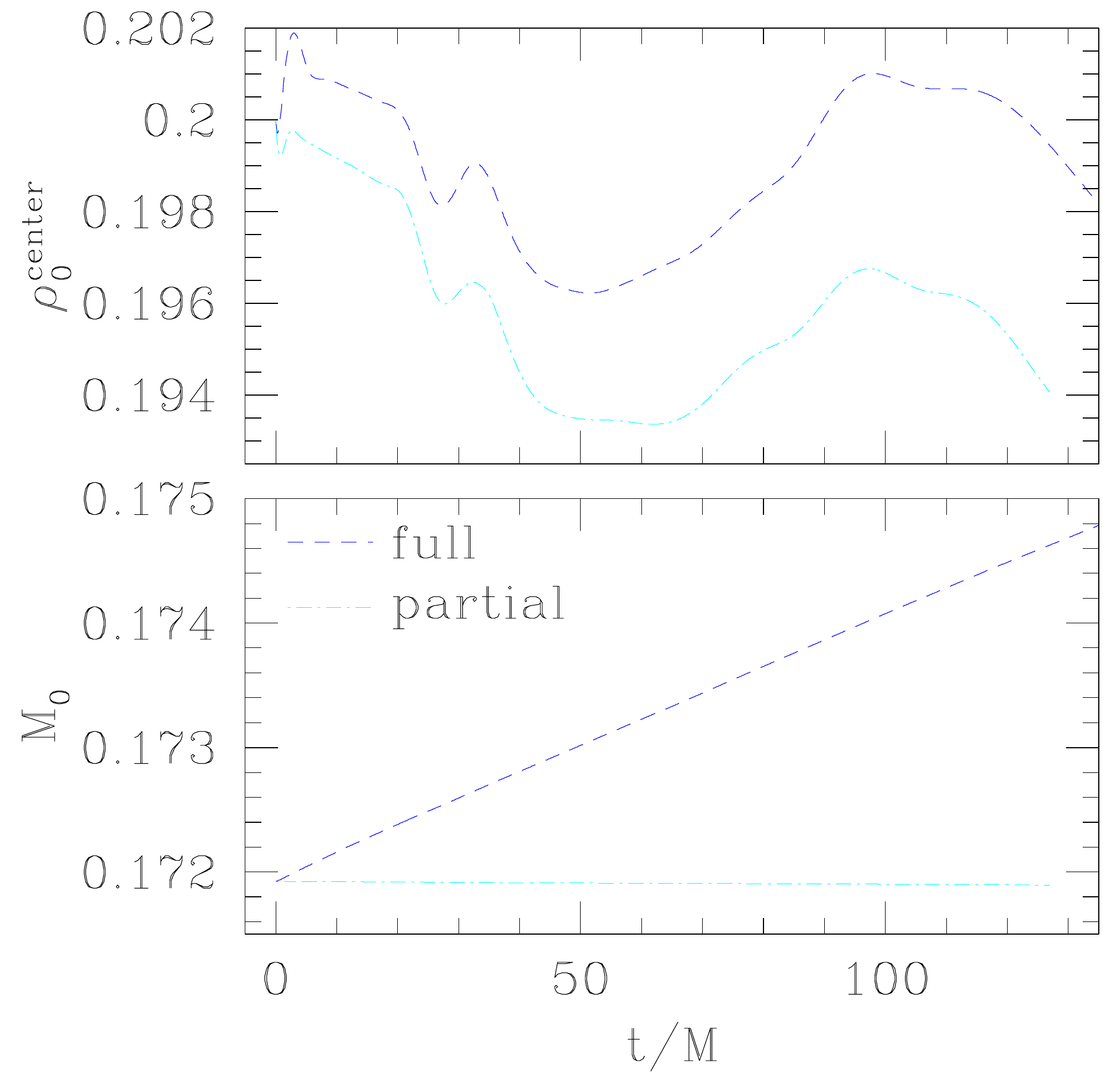}
\end{center}
\caption{A comparison between the full and partial approaches, for the resolution $N=3/2$ shown in Figs.~\ref{Fig:TOV_density} and \ref{Fig:TOV_mass}.  The top panel shows that the errors in the central density are somewhat smaller in the full approach than in the partial approach.  The bottom panel shows that, as expected, the rest mass in the partial approach is conserved almost exactly; the only error in the rest-mass arises from the treatment of the atmosphere (see Sects.~\ref{subsubsec:hydro_reference} and \ref{subsubsec:hydro_numerics}.)}
\label{Fig:TOV_comparison}
\end{figure}

As a last example we consider a static equilibrium star, i.e.~a solution to the Tolman-Oppenheimer-Volkoff (TOV) equations \cite{Tol39,OppV39}.  We set up the initial data as a relativistic polytrope with $\Gamma = 2$ (see Section \ref{subsubsec:eos}).   For this polytropic index, the maximum-mass model has a central rest-mass density of $\rho_0^{\rm max} = 0.318$, a rest-mass of $M_0 = 0.180$, and an ADM mass of $M = 0.164$ (in our code units with $\kappa = 1$).   For the simulations shown in this Section we choose a star with central density of $\rho_0^{\rm max} = 0.2$, for which the rest mass is $M_0 = 0.172$ and the ADM mass is $M = 0.157$.   The areal radius of this star is $R = 0.866 = 5.52 M$, while the isotropic radius is $r = 0.700 = 4.46 M$.   We evolve this star with non-advective 1+log slicing (\ref{1+log_non_ad}) and zero shift, so that, analytically, the star should remain static.  Any deviation from the initial data is therefore a measure of the numerical error.

We place the center of this spherical star at $r = 0.5 = 3.18M$ on the northern axis ($\theta = 0$), so that the origin of the coordinate system is inside the star (see Fig.~\ref{Fig:TOV}).    We evolve this stellar model on a grid that extends to an outer boundary at $r_{\rm max} = 3 = 19.1 M$.  We use (\ref{ang_res}) to match the angular resolution to the radial solution at the center of the star; in the following we show results for grid resolutions of $(N64,N32,2)$ with $N = 1$, $3/2$ and $2$.

The colored surface in Fig.~\ref{Fig:TOV} shows a slice of the rest-mass density $\rho_0$ at time $t = 175.4 M$, evolved on the highest-resolution grid with $N = 2$.  The contour lines demonstrate that the star remains nearly spherical.  Also included in the figure is a wireframe that shows the initial data. It can barely be noticed in the figure that the star has slightly drifted towards the origin -- a numerical artifact very similar to that described in Section \ref{subsubsec:Schwarzschild} for a single black hole.  Again, this effect becomes smaller with increasing resolution.  The origin of the coordinate system can be seen in the envelope of the star; clearly, the presence of the coordinate singularities at the origin and on the axis do not cause any problems in this simulation.

In Fig.~\ref{Fig:TOV_density} we show $\rho_0^{\rm center}$, the rest-mass density $\rho_0$ interpolated to the center of the star at $r = 0.5 = 3.18 M$ and $\theta = 0$.    Analytically, this value should remain at $\rho_0^{\rm max} = 0.2$, but numerical error causes small deviations.  In the top panel of Fig~\ref{Fig:TOV_density} we show $\rho_0^{\rm center}$ for the three different resolution $N = 1$, $3/2$ and $2$ as a function of time, in the bottom panel we show the relative errors $(\rho_0^{\rm center} - \rho_0^{\rm max})/\rho_0^{\rm max}$ multiplied with factors $N^2$.   Our results show second-order convergence until a time of $t \sim 25 M$; after that, the center of the star is affected by numerical errors originating from the stellar surface, where discontinuous derivatives of fluid and metric variables lead to a slower rate of convergence.

In Fig.~\ref{Fig:TOV_mass} we show similar results for the total rest mass $M_0$, given by the integral (\ref{restmass}).  In the top panel we show numerical values of $M_0$ as a function of time for the three different resolutions $N = 1$, 3/2 and 2; in the bottom panel we show the relative numerical errors multiplied with $N^3$.  These findings suggest approximately third-order convergence of the rest mass.  As discussed in Sect.~\ref{subsubsec:hydro_reference}, the origin of this error is the new term on the right-hand side of eq.~(\ref{continuity_ref_ex}), which appears in the ``full" reference-metric formulation of hydrodynamics.  In Fig.~\ref{Fig:TOV_comparison} we therefore compare results for both the rest-mass density and total rest-mass as obtained in the ``full" and the ``partial" approaches.  The top panel of Fig.~\ref{Fig:TOV_comparison} shows that the numerical errors for the rest-mass density are somewhat smaller in the full approach than in the partial approach, while the bottom panel demonstrates that, as expected, the total rest mass is conserved much better in the partial approach (the only error originates from our treatment of the atmosphere, see Section \ref{subsubsec:hydro_numerics}).  

\section{Summary and Discussion}
\label{sec:summary}

We have recently developed a new approach for numerical relativity simulations in spherical polar coordinates \cite{BauMCM13,MonBM14}.  Our approach is not based on a regularization of the equations, and instead deals with the coordinate singularities at the origin and on the axis of the coordinate system by using a reference-metric formulation of the equations, a factoring of all tensor components, and a partially implicit Runge-Kutta method.  Unlike previous approaches that employed a regularization of the equations, our method does not rely on any symmetry assumptions.  

While we have previously presented several tests of these methods, most of these test cases featured a symmetry about the origin.  In this paper we therefore present new ``off-centered" simulations that highlight the stability of our methods in the absence of such a symmetry.  We perform several stringent tests to show that the coordinate singularity at the origin of the coordinate system does not pose any computational problems -- even for a black hole that drifts through the origin.

The other purpose of this paper is to discuss an alternative implementation of the reference-metric formulation of relativistic hydrodynamics.   Unlike in \cite{MonBM14}, we here apply the same factoring of hydrodynamical tensor components as we use for Einstein's field equations.  We perform several simulations to test and calibrate this implementation.  Perhaps most importantly, we demonstrate that our method is able to follow the propagation of a shock wave through the origin of the spherical polar coordinate system. We are not aware of a previous solution to this problem.

We also discuss the respective advantages of the ``full" and ``partial" approach in our reference-metric formulation of relativistic hydrodynamics (see Sect.~\ref{subsubsec:hydro_reference}).   The partial approach applies this formulation only to the Euler equation, while the full approach applies it to all hydrodynamical variables and equations.  The advantage of the partial approach is that the right-hand side of the continuity equation vanishes, so that, in HRSC implementations, the rest mass is conserved exactly (except for errors originating from the treatment of the atmosphere).  The full approach, on the other hand, allows for a more accurate treatment of the origin of the coordinate system, and avoids any spiky or singular behavior even for a shock-wave passing through the origin.  Which one of the two approaches is preferable may therefore depend on the application.

\acknowledgements

 We would like to thank Reiner Birkl, Scott Noble and Jeffrey Winicour for asking us about our code's ability to simulate the propagation of a shock wave through the origin of the spherical polar grid, and thereby prompting this study.  Both TWB and PJM would like to thank for hospitality -- TWB at the Max-Planck-Institute f\"ur Astrophysik and PJM at Bowdoin College.  This work was supported in part by NSF grants PHY-1063240 and PHY-1402780 to Bowdoin College, and the Deutsche Forschungsgemeinschaft (DFG) through its Transregional Center SFB/TR7 ``Gravitational Wave Astronomy''.

%

\end{document}